\documentclass[12pt]{article}

\usepackage[utf8]{inputenc}
\usepackage[lighttt]{lmodern}
\usepackage{amsmath,amsfonts,amssymb}
\usepackage{mathtools, nccmath}
\usepackage{bm}
\usepackage{comment}
\usepackage{graphicx}
\usepackage{fancyvrb}
\usepackage[left=2.50cm, right=2.50cm, top=2.50cm, bottom=2.50cm]{geometry}
\usepackage{hyperref}
\usepackage[normalem]{ulem}
\usepackage{color}
\usepackage{setspace}
\usepackage[backend=biber,natbib=true,maxbibnames=9, sorting=none,url=false,date=year,doi=false,giveninits=true,isbn=false,bibwarn=true]{biblatex}

\newcommand{\gd}{Gr\"obner deformation~}

\definecolor{boxcolor}{RGB}{235,245,255}
\newcommand{\mybox}[1]{\begin{center}\fcolorbox{black}{boxcolor}{\parbox[c]{16.5cm}{#1}}\end{center}}
\newcommand{\inbox}[2]{\begin{tabular}{rl}{\tt In[#1]:= } #2\end{tabular}}
\newcommand{\outbox}[2]{\begin{tabular}{rl}{Prints $\Rightarrow$\hspace{0.85mm}} #2\end{tabular}}
\newcommand{\boxsplit}{\rule{16.5cm}{0.5pt}}
\newcommand{\gibspace}[2]{\begin{tabular}{rl}{\tt \hphantom{aaaaaaa} } #2\end{tabular}}

\begin{document}

\begin{center}

{\Large\bf \texttt{FeynGKZ}: a \textit{Mathematica} package for solving Feynman integrals using GKZ hypergeometric systems\\}
 
 \medskip

\vspace{1cm}
    {\bf{B. Ananthanarayan}$^{*}$, \bf Sumit Banik$^{\dagger}$, Souvik Bera$^{\ddagger}$, Sudeepan Datta$^{\star}$}\\[0.3cm]
{\small  Centre for High Energy Physics, Indian Institute of Science, \\	Bangalore-560012, Karnataka, India}\\[0.5cm]
\end{center}

\begin{abstract}
In the Lee-Pomeransky representation, Feynman integrals can be identified as a subset of Euler-Mellin integrals, which are known to satisfy Gel\textsc{\char13}fand-Kapranov-Zelevinsky (GKZ) system of partial differential equations. Here we present an automated package to derive the associated GKZ system for a given Feynman diagram and solve it in terms of hypergeometric functions using two equivalent algorithms, namely the triangulation method and the Gr\"obner deformation method. We present our code in the form of a \textit{Mathematica} package \texttt{FeynGKZ.wl} which requires the softwares \texttt{polymake}, \texttt{Macaulay2} and \texttt{TOPCOM}, and the packages \texttt{AMBRE} and \texttt{Olsson.wl} as dependencies. As applications of the package, we find series solutions to the GKZ systems of several one-loop and two-loop Feynman integrals. These are included in the file \texttt{Examples.nb} that can be downloaded along with the package from \href{https://github.com/anant-group/FeynGKZ}{\texttt{GitHub}}.
\end{abstract}
\vspace{9cm}

\small{$*$ anant@iisc.ac.in }

\small{$\dagger$ sumitbanik@iisc.ac.in }

\small{$\ddagger$ souvikbera@iisc.ac.in}

\small{$\star$ sudeepand@iisc.ac.in}

\newpage
\textbf{Program summary}:
\begin{itemize}
    \item \textit{Program Title} : \texttt{FeynGKZ.wl}, version 1.0
    \item \textit{Developer's repository link}: \href{https://github.com/anant-group/FeynGKZ}{https://github.com/anant-group/FeynGKZ}
    \item \textit{Licensing provisions}:  GNU General Public License 3 (GPL)
    \item \textit{Programming language} : \texttt{Wolfram} \textit{Mathematica} version 13.0 or higher
    \item \textit{External routines/libraries used}: \texttt{Macaulay2} version 1.20, \texttt{TOPCOM} version 0.17.8, \texttt{polymake} version 4.6, \texttt{AMBRE} version 2.1.1  and \texttt{Olsson.wl}
    \item \textit{Nature of problem} : Deriving the GKZ system associated with a given Feynman integral, and solving it in terms of multivariate hypergeometric functions. 
    \item \textit{Solution method }: Automating the triangulation and Gr\"obner deformation methods for obtaining $\Gamma$-series solutions to the GKZ hypergeometric system associated with a given Feynman integral.
    \item \textit{References}:
    \begin{enumerate}
        \item \href{http://www.wolfram.com/mathematica/}{\texttt{Wolfram} \textit{Mathematica}}; proprietary software.
        \item \href{https://polymake.org/}{\texttt{polymake}}, open-source software;
        \item \href{http://www2.macaulay2.com/Macaulay2/}{\texttt{Macaulay2}}, open-source software;
        \item \href{https://www.wm.uni-bayreuth.de/de/team/rambau_joerg/TOPCOM/index.html}{\texttt{TOPCOM}}, open-source software;
        \item \href{https://jgluza.us.edu.pl/ambre/}{\texttt{AMBREv2.1.1.m}}, open-source \textit{Mathematica} package;
        \item \href{https://arxiv.org/abs/2201.01189}{\texttt{Olsson.wl}}, open-source \textit{Mathematica} package.
    \end{enumerate}
\end{itemize}

\newpage
\section{Introduction}

The evaluation of multi-loop multi-scale Feynman integrals at ever increasing levels of complexity, has been of great interest and importance as they are the cornerstone of high precision theoretical calculations to match the precision level of present and future colliders.
In particular, analytical results for these are of great importance, even if in special cases, as they
would offer interesting insights as well as useful cross-checks with the corresponding numerical results. For an overview on the different analytical and numerical techniques, see \cite{Smirnov:book} and \cite{Weinzierl:2022eaz}.

In dimensional regularization, Feynman integrals can be expressed in terms of multiple hypergeometric functions \cite{Srivastava:1985} which are widely known objects in mathematics literature.
Several methods have been developed to solve Feynman integrals in terms of hypergeometric functions. One of the most popular being the Mellin-Barnes (MB) method \cite{Smirnov:1999gc,Tausk:1999vh,Ananthanarayan:2020fhl}. Here we focus on another powerful method which is based on the work of Gel\textsc{\char13}fand, Kapranov, and Zelevinsky (GKZ) which explores a system of partial differential equations (PDEs) whose solutions are hypergeometric functions and Euler-Mellin (EM) integrals.
While studying the differential equations satisfied by Feynman integrals,
M. Kalmykov and B. Kniehl were among the first to point out \cite{Kalmykov:2012rr}, based on the work of F. Beukers \cite{Beukers2011MonodromyOA}, that Feynman integrals could in all likelihood be solutions of an underlying GKZ system.

After the discovery of the
Lee-Pomeransky (LP) representation \cite{Lee:2013hzt} of the Feynman integral, the connection between Feynman integrals and GKZ is more straightforward. This has been explicitly shown in \cite{delaCruz:2019skx,Klausen:2019hrg}, where the associated $\mathcal{A}$-matrix of the GKZ system can be constructed from the $G$-polynomial of the LP representation. The GKZ system can also be derived from the PDE satisfied by the MB representation of the Feynman integral as discussed in \cite{Feng:2019bdx,Feng:2022kgh,Feng:2022ude}. The derived GKZ system can then be used to study different analytic properties of the original Feynman integral. For example, one can derive series solutions using two alternative algorithms as discussed in \cite{delaCruz:2019skx} and \cite{Klausen:2019hrg}. Alternatively, the GKZ framework is also useful to study the singular locus of Feynman integrals \cite{Klausen:2021yrt}, vector spaces generated by Feynman integrals \cite{Agostini:2022cgv}, Cohen-Macaulay property of Feynman integrals \cite{Tellander:2021xdz}, matroids attached to Feynman integrals \cite{Walther:2022pli}, etc. It is also noted that the PDEs of the GKZ system is closely related to the PDEs appearing in the Yangian bootstrap approach \cite{Loebbert:2019vcj}. Furthermore, due to the connection between GKZ systems and $D$-module theory, one can efficiently derive first-order differential equations \cite{Chestnov:2022alh} satisfied by a basis of master integrals of a Feynman integral. The realisation of parametric Feynman integrals as EM integrals has also of late provided a novel computational algorithm to numerically evaluate Euclidean Feynman integrals, employing methods inspired from tropical geometry \cite{Borinsky:2020rqs}.

Despite recent rapid progress in the theoretical side of these emerging topics, there are no publicly available automated computational tools for the study of GKZ system associated to Feynman integrals. We take the first step in this direction by presenting the \textit{Mathematica} package \texttt{FeynGKZ.wl} which can derive the GKZ system satisfied by a given Feynman integral.  Subsequently, we find the series solutions of the derived GKZ system using two equivalent algorithms. One of these relies on Gr\"obner deformations (GD) which loosely speaking is a generalization of the celebrated
Frobenius method for ODEs in a single variable, to PDEs in several variables \cite{SST}. The other technique relies on taking suitable triangulations of the polytope defined by a certain point configuration. 
It may be noted that both methods yield solutions in terms of multiple hypergeometric functions. \texttt{FeynGKZ.wl} uses \texttt{Macaulay2} and \texttt{TOPCOM} \cite{Rambau:TOPCOM:2002} for intermediate steps in the GD and triangulation method, respectively. It also uses \texttt{polymake} \cite{Gawrilow2000polymakeAF} to calculate the normalized volume of the Newton polytope associated to the $\mathcal{A}$-matrix of the GKZ system. The \textit{Mathematica} packages \texttt{AMBRE} \cite{Gluza:2007rt} and \texttt{Olsson.wl} \cite{Ananthanarayan:2021yar} are used as well to derive MB representation and to express double-sums in terms of known hypergeometric functions, respectively.

The outline of the paper is as follows. In section \ref{sec:GKZoverview}, we provide an overview of the GKZ hypergeometric systems. In section \ref{FP-LP}, we briefly recap scalar Feynman integrals and their LP representation. In section \ref{GKZfromLP}, we recall how to derive the GKZ system from the LP representation and find series solutions using either the GD or triangulation approach. In Section \ref{sec:manual}, we discuss the usage of the external modules of \texttt{FeynGKZ.wl}. In section \ref{sec:Example} we demonstrate the package by taking a simple one-loop massive Feynman integral. Finally, in Section \ref{sec:Conclusion} we present our conclusions and remarks. In appendix \hyperlink{ExtModules}{A} and \hyperlink{GlobalVars}{B} we provide the usage of some additional external modules and global variables of the package, respectively. In appendix \hyperlink{Examples}{C}, we list the Feynman integrals we have demonstrated in the file \texttt{Examples.nb} which can be downloaded along with the package \texttt{FeynGKZ.wl} from \href{https://github.com/anant-group/FeynGKZ}{https://github.com/anant-group/FeynGKZ}.

\section{Overview of GKZ hypergeometric systems} \label{sec:GKZoverview}

In this section, we provide a brief overview of the GKZ hypergeometric system. To keep the mathematical framework short, we provide only the notions required for the computational aspects. The full theory requires more rigorous mathematical ideas which are not required for our purpose. The interested reader will find those in the following \cite{SST,Cattani:2006,Stienstra:2005nr,Beukers:notes,Hibi}

\subsection{Setting up the GKZ system}\label{GKZsys}

Let $\mathcal{A}= a_{ij}$ be a $(n+1)\times N$ matrix with $n+1 \leq N$, having rank $n+1$ and all elements are integers. We assume that the vector $(1,1,\dots,1)$ lies in the $\mathbb{Q}$-row span of $\mathcal{A}$ and the column vectors of $\mathcal{A}$ generate $\mathbb{Z}^{n+1}$ over $\mathbb{Z}$. The  {toric ideal} associated with the matrix $\mathcal{A}$ is defined as,
\begin{align}
    I_\mathcal{A} := \langle \partial^u - \partial^v | \mathcal{A} u = \mathcal{A} v ~ ;~ u,v \in \mathbb{N}^N  \rangle \in \mathbb{C}[\partial]
\end{align}
where $\mathbb{C}[\partial] =  \mathbb{C}[\partial_1 , \dots ,\partial_N]$ is a commutative polynomial ring.

Given a column vector $\beta = (\beta_0 ,\dots , \beta _n)^T \in \mathbb{C}^{n+1}$, the  {GKZ hypergeometric system} characterised by matrix $\mathcal{A}$ and parameters $\beta$ is defined as the left ideal $H_\mathcal{A}(\beta)$ generated by $I_\mathcal{A}$ and $\langle \mathcal{A}\cdot\theta-\beta \rangle$. 
\begin{equation}\label{GKZ_PDE}
    H_\mathcal{A}(\beta)  = I_\mathcal{A} \cup \langle \mathcal{A}\cdot\theta-\beta \rangle \in \mathbb{C}[z_1,\dots, z_N,\partial_1,\dots,\partial_N]
\end{equation}
where the vector $\theta = (\theta_1,\dots,\theta_N)^T$ contains the Euler operators $\theta_i =  z_i \partial_i$ as its elements. This is also known as the $\mathcal{A}$-hypergeometric ideal due to the presence of the matrix $\mathcal{A}$.  A holomorphic function $\phi(z_1,\dots,z_N)$ is called the $\mathcal{A}$-hypergeometric function if $H_\mathcal{A}(\beta) \bullet \phi =0$. Here we are only interested in non-logarithmic series solutions of the above system. 

\subsection{EM integrals as solutions to GKZ hypergeometric systems}
It is pertinent here to discuss the solutions to GKZ systems in terms of EM integrals, for we shall later see that a generalized version of Feynman integrals could be understood in terms of such solutions. 

To define the EM integral, we consider a generic Laurent polynomial:
\begin{equation}\label{eqn:Laurentpoly}
    f(\alpha_1,...,\alpha_n;z_1,...,z_N) \equiv f_z(\alpha):=\sum_{j=1}^N z_j \prod_{i=1}^n  \alpha_i^{a_{ij}}; \quad a_{ij} \in \mathbb{Z}
\end{equation}

We can now define an EM integral as a generalization of the Mellin transform of the rational function $1/f$:
\begin{equation}\label{eqn:EMInt}
    I_{f_z}(\nu,\nu_0)=\int_{\mathbb{R}_n^+} {d^n\alpha}\, \alpha^{\nu-1}f_z(\alpha)^{-\nu_0}
\end{equation}
where, $d^n\alpha=\prod_{i=1}^nd\alpha_i$, and $\alpha^{\nu-1}=\prod_{i=1}^n\alpha_i^{\nu_i-1}$.

Analytic and convergence properties of such integrals have been detailed in \cite{EMIntegrals,delaCruz:2019skx,Klausen:2019hrg} by considering the Newton polytope corresponding to the Laurent polynomial, namely, in Theorems 2.2 and 2.4 in \cite{EMIntegrals}.

It has been shown in \cite{EMIntegrals} (Theorem 4.2) that
for generic $z_j$, the EM integral $I_{f_z}$ with parameters $(\nu,\nu_0)=(\nu_1,...,\nu_n,\nu_0)$ could be understood as a solution to a GKZ hypergeometric system of the form Eq.\eqref{GKZ_PDE} with the following data:
\begin{equation}\label{eqn:EMIntsolforGKZ}
\begin{split}
    \mathcal{A} & = \{a_{ij}; i \in \{1,...,n+1\}, j \in \{1,...,N\}\}|a_{ij}=1; i=1\} \\
    \beta & = -(\nu_0,\nu)^T=-\underline{\nu}
\end{split}
\end{equation}
where, $\underline{\nu}=(\nu_0,\nu)^T$.

Thus, the GKZ system associated with $f_z(\alpha)$ takes the following form in terms of the parameter vector $\underline{\nu}$:
\begin{equation}
    H_\mathcal{A}(\underline{\nu})  = I_\mathcal{A} \cup \langle \mathcal{A}\cdot\theta+\underline{\nu} \rangle
\end{equation}

\section{Scalar Feynman integrals and the Lee-Pomeransky representation} \label{FP-LP}
In the momentum-space representation, scalar Feynman integrals take the following form:
\begin{equation}
    I_\Gamma(\nu,D)=\int\prod_{r=1}^l \frac{d^Dk_r}{i\pi^\frac{D}{2}}\frac{1}{\prod_{j=1}^{n}(-q_j^2+m_j^2)^{\nu_j}}
\end{equation}
where, $l$ is the number of loops, $D$ is the space-time dimension, $\nu=(\nu_1,...,\nu_n)$ are the propagator powers. $k_r$-s and $q_j$-s are the loop-momenta and internal-momenta for the Feynman graph $\Gamma$.

Introducing the \textit{Feynman parameters} $\alpha=\{\alpha_i:i \in \{1,...,n\}\}$ for all the internal edges, and using the notation \(\omega=\nu-\frac{lD}{2}\), a parametric representation for the integral above is obtained:

\begin{equation}
    I_\Gamma(\nu,D)=\Gamma(\omega)\Big(\prod_{i=1}^{n}\int_{\alpha_i=0}^1 \frac{d\alpha_i\, \alpha_i^{\nu_i-1}}{\Gamma(\nu_i)}\Big)\frac{\delta(1-\sum_{j=1}^{n}\alpha_j)}{F(\alpha)^{\omega} U(\alpha)^{\frac{D}{2}-\omega}}
\end{equation}
This expression is dubbed the \textit{Feynman parametric representation} for the integral. $U(\alpha)$ and $F(\alpha)$ are called the first and second Symanzik polynomials, respectively.

There is another alternative representation which involves a single polynomial  $G(\alpha)=U(\alpha)+F(\alpha)$,

\begin{equation}\label{eq:LP}
\begin{split}
    I_\Gamma(\nu,D) & = \frac{\Gamma(\frac{D}{2})}{\Gamma(\frac{D}{2}-\omega)} \Big(\prod_{i=1}^{n}\int_{\alpha_i=0}^\infty \frac{d\alpha_i\, \alpha_i^{\nu_i-1}}{\Gamma(\nu_i)}\Big) G(\alpha)^{-\frac{d}{2}} \\
    & = \frac{\Gamma(\frac{D}{2})}{\Gamma(\frac{D}{2}-\omega)\Gamma(\nu)} \int_{\mathbb{R}^n_+} d\alpha\, \alpha^{\nu-1} G(\alpha)^{-\frac{d}{2}}
\end{split}
\end{equation}
where, we use the following notation:
\begin{equation}
\begin{split}
    d\alpha  = \prod_{i=1}^{n}d\alpha_i\hspace{0.3cm}, \hspace{0.7cm}
   \alpha^{\nu-1}  = \prod_{i=1}^{n}\alpha_i^{\nu_i-1} \hspace{0.3cm}, \hspace{0.7cm}
   \Gamma(\nu)  = \prod_{i=1}^{n}\Gamma(\nu_i)
\end{split}
\end{equation}
This representation is called the \textit{Lee-Pomeransky representation} \cite{Lee:2013hzt} for the Feynman integral, and the polynomial $G(\alpha)$ is called the \textit{Lee-Pomeransky polynomial}. Hereafter, we shall call this polynomial the $G$-polynomial.

For further details on the different representations of Feynman integrals, we refer the interested reader to \cite{Weinzierl:2022eaz}.

\section{Solution methodology}\label{GKZfromLP}

\subsection{The toric $G$-polynomial and the $\mathcal{A}$-matrix}\label{AmatToPointConf}


Polynomials with generic coefficients {of its monomials} are called \textit{toric polynomials}. It has been shown in \cite{delaCruz:2019skx} that a toric version of the $G$-polynomial defines a GKZ-system. This toric version is expressed as follows:
\begin{equation}\label{eqn:toricG}
    G_z(\alpha)=\sum_{a_j \in A}z_j \alpha^{a_j}=\sum_{j=1}^N z_j \prod_{i=1}^n \alpha_i^{a_{ij}}
\end{equation}
where, $N$ is the number of monomials, $z_j$ are generic coefficients that contain information about the kinematics in the non-generic limit, and are complex numbers  $z_j \in (\mathbb{C}\backslash{0})^N$. $A$ is a finite set with $N$ pairwise distinct column vectors $a_j =(a_{1j},...,a_{nj})^T \in \mathbb{Z}^n_{\geq0}$. As such, $A$ is a matrix of order $n \times N$.

The matrix $A$ defines a configuration of $N$ points in affine space $\mathbb{Z}^n$. The convex hull of these points defines a convex polytope:
\begin{equation}
    P=\text{Conv}(A):=\Big\{\sum_{j=1}^Nk_ja_j \Big|k \in \mathbb{R}^N_{\geq0},\sum_{j=1}^Nk_j=1 \Big\}
\end{equation}

The toric $G$-polynomial is a special case of a Laurent polynomial where only non-negative exponents appear. Therefore by Eq.\eqref{eqn:EMIntsolforGKZ}, we can write the associated $\mathcal{A}$-matrix as follows:
\begin{equation}
    \mathcal{A}=\begin{pmatrix}
                   1\\
                   A
                \end{pmatrix}=\begin{pmatrix}
                                 1 & 1 & ... & 1\\
                                 a_1 & a_2 & ... & a_N
                              \end{pmatrix} \in \mathbb{Z}_{\geq0}^{(n+1)\times N}
\end{equation}
where, $a_i \in \mathbb{Z}^n_{\geq 0}$.

This condition ensures that the columns in $\mathcal{A}$ are the homogeneous coordinates of the point configuration defined by $A$. This is the convention used in \texttt{TOPCOM} \cite{Rambau:TOPCOM:2002}. More details about this homogeneity condition can be found in \cite{TriangulationsBook}.

The Newton polytope associated to the toric $G$-polynomial is defined as the convex hull of the exponent vectors:
\begin{equation}
    \Delta_{G_z}:=\text{Conv}(\{a_j=(a_{1j},...,a_{nj})^T|z_j \neq 0\},j \in \{1,...,N\})
\end{equation}
For details about convex polytopes in general and the Newton polytope in particular, we refer the reader to \cite{Klausen:2019hrg,Weinzierl:2022eaz}.

From this discussion, we can see that $\Delta_{G_z}=\text{Conv}(A)$.

\subsection{The triangulation method}

\subsubsection{Triangulations of $\Delta_{G_z}$}

A triangulation of $\Delta_{G_z}$ is its subdivision into simplices $\{\sigma_1,...,\sigma_r\}$, such that union of all simplices gives the full polytope, and the intersection of two distinct simplices is either empty or a proper face of both. Here, $\sigma_i \subset \{1,...,N\}$ is an index set, $i \in \{1,...,r\}$, and $\overline{\sigma}=\{1,...,N\}\backslash{\sigma}$ is the complement of $\sigma$.

\textit{Regular} or \textit{coherent} triangulations $T(h)=\{\sigma_1,...,\sigma_r\}$ are those for which there exists a height vector $h \in \mathbb{R}^N$, such that for every simplex $\sigma_i$ of such a triangulation, there exists another vector $p_i \in \mathbb{R}^{n+1}$ satisfying
\begin{equation}
    \begin{split}
        p_i \cdot a_j & = h_j \qquad \text{for} \qquad j \in \sigma_i \\
        p_i \cdot a_j & < h_j \qquad \text{for} \qquad j \notin \sigma_i
    \end{split}
\end{equation}
It has been shown in \cite{TriangulationsBook} that a regular triangulation can always be obtained, given an arbitrary convex polytope. 

As mentioned earlier in section \ref{GKZsys}, we consider the rank of $\mathcal{A} \in \mathbb{Z}^{(n+1) \times N}$ to be $n+1$. Let $\delta$ be the greatest common divisor of all $(n+1)\times(n+1)$ minors of $\mathcal{A}$. Then the \textit{normalized volume} of $\sigma$ is given by $\text{vol}_0(\sigma)=|\text{det} \mathcal{A}_\sigma|/\delta$ where, $\mathcal{A}_\sigma$ denotes the submatrix formed by retaining the columns indexed by $\sigma$ and dropping the others. This definition could be found in \cite{Hibi,DBLP:books/daglib/0030216}.

\textit{Unimodular} triangulations $\{\sigma_1,...,\sigma_r\}$ are those for which, $\text{vol}_0(\sigma_i)=1 \: \forall \: i \in \{1,...,r\}$.

\subsubsection{Generalized Feynman integrals}

Employing the notion of the toric $G$-polynomial with generic coefficients $z_j \in (\mathbb{C}\backslash{0})^N$ such that $Re(z_j)>0$, we can consider a generalized Feynman integral as the meromorphic continuation of the integral 
\begin{equation}\label{eqn:genFI}
    I_{G_z}(\nu,\nu_0)=\Gamma(\nu_0)\int_{\mathbb{R}_+^n}d\alpha\, \alpha^{\nu-1}G_z(\alpha)^{-\nu_0}
\end{equation}
defined on $\underline{\nu}=(\nu_0,\nu) \in \mathbb{C}^{n+1}$, with $\nu_0=\frac{D}{2}$. 

From the discussion in \cite{Klausen:2019hrg}, another representation of $I_{G_z}(\nu,\nu_0)$ could be obtained as a multi-fold MB integral:
\begin{equation}
    I_{G_z}(\nu,\nu_0)=\frac{z_\sigma^{-\mathcal{A}_\sigma^{-1}\underline{\nu}}}{\text{vol}_0(\sigma)} \int_{\gamma} \frac{dt}{(2\pi i)^r}\Gamma(t)\Gamma(\mathcal{A}_\sigma^{-1}\underline{\nu}-\mathcal{A}_\sigma^{-1}\mathcal{A}_{\overline{\sigma}}t)z^{-t}_{\overline{\sigma}}z^{\mathcal{A}_\sigma^{-1}\mathcal{A}_{\overline{\sigma}}t}_\sigma
\end{equation}

From the above, a special case follows where $N=n+1$. If there exists some $\text{ region } R \subseteq \mathbb{C}^{n+1}$ such that $ I_{G_z}(\nu,\nu_0)$ is absolutely convergent for $\underline{\nu}\in R$, then $\text{det }\mathcal{A}_{\sigma} \neq 0$, and the Feynman integral takes a simple form:
\begin{equation}\label{n=N+1}
    I_{G_z}(\nu,\nu_0)=\frac{\Gamma(\mathcal{A}_{\sigma}^{-1}\underline{\nu})}{\text{vol}_0(\sigma)}z^{-\mathcal{A}_{\sigma}^{-1}\underline{\nu}}
\end{equation}
where, owing to the square matrix form, $\sigma$ contains all the columns, that is, $\mathcal{A}_{\sigma}=\mathcal{A}$. Thus, $\text{vol}_0(\sigma)=|\text{det }\mathcal{A}|/\delta$.

\subsubsection{Generalized Feynman integrals as GKZ hypergeometric functions}
Comparing Eq.\eqref{eqn:toricG} with Eq.\eqref{eqn:Laurentpoly} and Eq.\eqref{eqn:genFI} with Eq.\eqref{eqn:EMInt} , we can see that generalized Feynman integrals could be understood as EM integrals. Thus, $I_{G_z}(\nu,\nu_0)$ satisfies the GKZ hypergeometric system $H_{\mathcal{A}}(\underline{\nu})$ with the associated data as specified in \eqref{eqn:EMIntsolforGKZ}, where $\nu_0=\frac{d}{2}$. For a formal proof, we refer the reader to \cite{Klausen:2019hrg} (Theorem 3.1).

A basis for the solution space of $H_{\mathcal{A}}(\underline{\nu})$ could be obtained using the regular triangulations of $\Delta_{G_z}$, with the restriction that the parameter vector $\underline{\nu}$ be very generic. Details of this procedure could be found in \cite{Gelfand1991HYPERGEOMETRICFT} (Theorem 2.17) and \cite{Gelfand1992GeneralHS}. 

Elements of this basis are called the $\mathit{\Gamma}$\textit{-series}. The number of such elements is given by $\text{rank } H_{\mathcal{A}}(\underline{\nu})=\text{vol}_0(\text{Conv}(A))$. Since GKZ hypergeometric systems are holonomic, it is ensured that the dimensionality of the solution space (and hence the number of such basis elements) is finite.

For unimodular regular triangulations, each simplex of the triangulation can lead to only one such series, and hence, we can label the $\Gamma$-series by their corresponding simplices. It can be shown that in the unimodular case, the $\Gamma$-series associated to some simplex $\sigma \in T$ takes the following form:
\begin{equation}\label{penult}
    \Phi_{\sigma}(\underline{\nu},z)=z_{\sigma}^{-\mathcal{A}_{\sigma}^{-1}\underline{\nu}}\sum_{\lambda\in\mathbb{N}_0^{|\overline{\sigma}|}}\frac{(\mathcal{A}_{\sigma}^{-1}\underline{\nu})_{\mathcal{A}_{\sigma}^{-1}\mathcal{A}_{\overline{\sigma}}\lambda}}{\lambda!}\frac{z_{\overline{\sigma}}^{\lambda}}{(-z_{\sigma})^{\mathcal{A}_{\sigma}^{-1}\mathcal{A}_{\overline{\sigma}}\lambda}}
\end{equation}

Using the Knudsen-Mumford theorem \cite{Knudsen1973,bruns2009polytopes}, it could be argued \cite{Klausen:2019hrg} that by adding monomials to the toric $G$-polynomial such that $G_z \rightarrow G'_{z'}$ (or equivalently, by scaling the underlying Newton polytope $\Delta_{G_z}$ with some integer scale-factor such that $\Delta_{G_z} \rightarrow \Delta_{G'_{z'}}$), an unimodular triangulation can always be obtained. As a result, it would be enough to consider only those regular triangulations of $\Delta_{G_z}$ that happen to be unimodular as well, in order to construct a basis for the solution space of $H_{\mathcal{A}}(\underline{\nu})$.

Starting with the case $\underline{\nu} \in (\mathbb{C}\backslash{\mathbb{Z}})^{n+1}$, we can write the corresponding generalized Feynman integral as a linear combination of the $\Gamma$-series:
\begin{equation}
    I_{G_z}(\nu,\nu_0)=\sum_{\sigma \in T} P_{\sigma}(\underline{\nu}) \Phi_{\sigma}(\underline{\nu},z)
\end{equation}
for $z_j \in \mathbb{C}^N$ such that $Re(z_j) \geq 0$ wherever the $\Gamma$-series converge. The prefactors $P_{\sigma}(\underline{\nu})$ are meromorphic functions in the parameter vector, and are to be computed in order to obtain a series representation for the Feynman integral.

One simple way to compute these prefactors is to exploit subtriangulations of the triangulation $T$. By a series of arguments, the authors in \cite{Klausen:2019hrg} show that one such trivial subtriangulation could be considered to be one of the simplices of $T$. In this case, using Eq.\eqref{n=N+1}, it could be shown that these prefactors are given by $P_{\sigma}(\underline{\nu})=\Gamma(\mathcal{A}_{\sigma}^{-1}\underline{\nu})$, since due to the unimodularity condition, $\text{vol}_0(\sigma)=1$.

As a result, using the above and Eq.\eqref{penult}, we can write the generalized Feynman integral as follows:
\begin{equation}\label{ult}
\begin{split}
    I_{G_z}(\nu,\nu_0) & = \Gamma(\mathcal{A}_{\sigma}^{-1}\underline{\nu})\sum_{\sigma\in T}z_{\sigma}^{-\mathcal{A}_{\sigma}^{-1}\underline{\nu}}\sum_{\lambda\in \mathbb{N}_0^{|\overline{\sigma}|}} \frac{(\mathcal{A}_{\sigma}^{-1}\underline{\nu})_{\mathcal{A}_{\sigma}^{-1}\mathcal{A}_{\overline{\sigma}}\lambda}}{\lambda!}\frac{z_{\overline{\sigma}}^{\lambda}}{(-z_{\sigma})^{\mathcal{A}_{\sigma}^{-1}\mathcal{A}_{\overline{\sigma}}\lambda}} \\
    \implies I_{G_z}(\nu,\nu_0) & = \sum_{\sigma\in T}z_{\sigma}^{-\mathcal{A}_{\sigma}^{-1}\underline{\nu}}\sum_{\lambda\in \mathbb{N}_0^{|\overline{\sigma}|}} \frac{\Gamma(\mathcal{A}_{\sigma}^{-1}\underline{\nu}+\mathcal{A}_{\sigma}^{-1}\mathcal{A}_{\overline{\sigma}}\lambda)}{\lambda!}\frac{z_{\overline{\sigma}}^{\lambda}}{(-z_{\sigma})^{\mathcal{A}_{\sigma}^{-1}\mathcal{A}_{\overline{\sigma}}\lambda}}
\end{split}
\end{equation}
where the series have a common convergence domain.

Eq.\eqref{n=N+1} and Eq.\eqref{ult} comprise the main equations that go into the geometric solution side of our package. Note that these equations provide us with solutions only for the generalized Feynman integral. Suitable limits for the variables $z_j$ need to be set in order to get back to the original Feynman integral, and the process has been discussed at length in \cite{Klausen:2019hrg}.

\subsection{The GD method}
Here we summarize the key steps of the GD method, which is based on the algorithm proposed by Saito, Sturmfels and Takayama (SST) \cite{SST}. Let $D_N$ be a Weyl algebra of $N$ variables over a field $\mathbb{K}$. It is generated by the variables $z_1,\dots,z_N, \partial_1,\dots,\partial_N$ where all the $z_i$'s commute among themselves and so do the $\partial_i$'s , but $[\partial_i , z_j] = \delta_{ij}$. An element $p$ that belongs to the Weyl algebra can be written in the canonical form, where all the $\partial_i$ are brought to the right by the use of commutation relations.
\begin{align}
    p = \sum_{i,j \in \mathbb{N}^N} c_{ij} z^i \partial^j 
\end{align}
Here $z^i$ for $i \in \mathbb{N}^N$ implies $z_1^{i_1} \dots z_N^{i_N}$ and the same for $\partial^j$. For a vector $w \in \mathbb{R}^N$,
\begin{align}
    \text{in}_{(-w,w)} (p) := \left\{ \sum_{r,s \in \mathbb{N}^N}  c_{rs} z^r \partial^s ~| -\langle w,r\rangle + \langle w,s\rangle \text{~is maximum}  \right\} 
\end{align}
where $\langle \_,\_ \rangle$ is the usual scalar product in $\mathbb{R}^N$. 
\\
For an ideal $I \subset D_N$, the  {Gr\"obner deformation} of $I$ with respect to the weight $w$ is defined as
\begin{align*}
     \text{in}_{(-w,w)} (I) := \langle  \text{in}_{(-w,w)} (p) | p \in I \rangle
\end{align*}
The  {initial ideal} is the \gd of the toric ideal $I_\mathcal{A}$ associated with matrix $\mathcal{A}$ with respect to a generic weight vector.

Let $M$ be a monomial ideal in $\mathbb{K}[\partial]$. A  {standard pair} associated with it is a pair $(\partial^a,\sigma)$ where $a\in \mathbb{N}^N$ and $\sigma \subset \{1,\dots, N\}$ satisfying 
\begin{itemize}
    \item $a_i = 0$ for all $i\in \sigma$
    \item for all choices of integers $b_j \geq 0$, $\partial^a \prod_{j\in\sigma }\partial_j^{b_j} \notin M$  \item for all $l \notin \sigma , \exists ~b_j \geq 0 : \partial^a \partial_l^{b_l} \prod_{j\in\sigma }\partial_j^{b_j} \in M$ 
\end{itemize}
Let $\mathcal{S}(M)$ be the set of all the standard pairs of $M$ then the decomposition of $M$ into irreducible monomial ideals can be found as
\begin{align*}
    M = \bigcap_{(\partial^a ,\sigma) \in \mathcal{S}(M)} \langle \partial_i^{a_i+1} : i \notin \sigma \rangle
\end{align*}
From the standard pair one can compute the  {indicial ideal} as \cite{Cattani:2006}
\begin{align*}
    \text{ind}_w (I_\mathcal{A}) =\bigcap_{\left(\partial^{a}, \sigma \right) \in \mathcal{S}\left(\operatorname{in}_{w}\left(I_{\mathcal{A}}\right)\right)}\left\langle\left(\theta_{j}-a_{j}\right), j \notin \sigma\right\rangle
\end{align*}
and {fake indicial ideal} as
\begin{align*}
    \text{find}_w (H_\mathcal{A}(\beta)) = \bigcap_{\left(\partial^{a}, \sigma\right) \in \mathcal{S}\left(\operatorname{in}_{w}\left(I_{\mathcal{A}}\right)\right)}\left\langle\left(\theta_{j}-a_{j}\right), j \notin \sigma\right\rangle+\langle \mathcal{A} \cdot \theta-\beta\rangle
\end{align*}
The roots of the fake indicial ideal are called  {fake exponents} of the GKZ system for a given weight. We denote the fake exponents by $v$.  
\\
Let us denote the kernel of the matrix $\mathcal{A}$ by
\begin{align}
    L := \text{ker}_\mathbb{Z}(\mathcal{A})
\end{align}
For $u \in L$, one can write $u = u_+ -u_-$, where $u_+$ and $u_-$ contains non-negative entries.
\\
Following \cite{SST}, we define the falling factorials for a vector $v\in \mathbb{K}^N$,
\begin{align*}
    \left[ v \right]_{u_-} &= \prod_{i:u_i<0} \prod_{j=1}^{-u_i} (v_i-j+1)\\
    \left[ u+ v\right]_{u_+} &= \prod_{i:u_i>0} \prod_{j=1}^{u_i}
    (u_i+j)
\end{align*}
In what follows, we assume that the vector $v$ does not contain negative integers so that
\begin{align*}
    \left[ u+ v\right]_{u_+} \neq 0\hspace{1cm} \text{for all~} u\in \mathbb{Z}^N 
\end{align*}
The series solution of the GKZ system can be written as \cite{SST}
\begin{align*}
    \phi_v := \sum_{u \in L} \frac{\left[v\right]_{u_-}}{ \left[ u+ v\right]_{u_+}} z^{u+v}
\end{align*}
This sums up the general strategy to find the canonical series solutions of a GKZ system $H_{\mathcal{A}}(\beta)$. 

\section{Documentation and Usage} \label{sec:manual}

In this section we describe the usage of the external modules of the package \texttt{FeynGKZ.wl}. There are a total of nine external modules that the user has access to. Here, we enlist only six of them that participate in creating the basic workflow of our package, namely, deriving the GKZ system for a given Feynman integral and solving the system in terms of hypergeometric functions. The remaining three modules with additional functionality have been described in appendix \hyperlink{ExtModules}{A}.

For a given Feynman integral, our first step is to find the $\mathcal{A}$-matrix associated to its GKZ system. This is done by using the \texttt{FindAMatrix} command. This command can accept two different input patterns of the Feynman integral: the momentum representation or the Schwinger parametric representation.
\mybox{
\textbf{\texttt{FindAMatrix[\{MomentumRep, LoopMomenta, InvariantList, Dim, Prefactor\}]}}
\\\\
or,
\\\\
\textbf{\texttt{FindAMatrix[\{\{U, F, PropagatorPowers\}, LoopNumber, InvariantList, Dim, Prefactor\}]}}
\\\\
Below, we provide details about the input arguments for \texttt{FindAMatrix}.
\begin{itemize}
    \item \texttt{MomentumRep}: A list containing information about all the propagators in the Feynman integral. It consists of multiple sub-lists, each of which has the following pattern: 
    \\\texttt{\{Propagator-Momentum, Propagator-Mass, Propagator-Power\}}.
    \item \texttt{U}: The first Symanzik polynomial corresponding to the given Feynman integral.
    \item \texttt{F}: The second Symanzik polynomial corresponding to the given Feynman integral.
     \item \texttt{PropagatorPowers}: A list containing all the powers of the propagators of the Feynman integral.
    \item \texttt{LoopMomenta}: A list containing the loop-momenta for the given Feynman integral.
    \item \texttt{LoopNumber}: The number of loops appearing in the given Feynman integral. It must be an positive integer.
    \item \texttt{InvariantList}: A list of kinematic substitutions.
    \item \texttt{Dim}: the space-time dimension.
    \item \texttt{Prefactor}: An overall factor that is independent of the loop-momenta.
    \item Options:
    \begin{itemize}
    \item \texttt{UseMB}: This option accepts a boolean value. Its default value is \texttt{False}. When set \texttt{True}, \texttt{FindAMatrix} calls \texttt{AMBREv2.1.1.m} to obtain the MB-representation corresponding to the given Feynman integral, and uses it to compute the $\mathcal{A}$-matrix.
    \end{itemize}
\end{itemize}
}
Upon a successful run with \texttt{UseMB $\rightarrow$ False}, \texttt{FindAMatrix} prints the \texttt{U}, \texttt{F} and \texttt{G} polynomials, the $\mathcal{A}$-matrix, the normalised Euclidean volume of the polytope defined by the $\mathcal{A}$-matrix as calculated by \texttt{polymake}, and the total time taken by the command to run.

The option \texttt{UseMB $\rightarrow$ True} works only when the Feynman integral is given in its momentum representation. When run with \texttt{UseMB $\rightarrow$ True}, \texttt{AMBRE} is called to derive the MB representation, which is subsequently used to derive the $\mathcal{A}$-matrix following an algorithm similar to \cite{Feng:2019bdx}. This option is still at an experimental stage as it works only when the derived MB representation is in a certain form. 

The next step is to compute either the set of all unimodular regular triangulations, or the set of all square-free initial ideals, for the GKZ system defined by the $\mathcal{A}$-matrix obtained in the previous step. One may do this by using either of the following two external modules:
\mybox{
\textbf{\texttt{FindTriangulations[FindAMatrixOut]}}
\\\\
This external module computes unimodular regular triangulations corresponding to the $\mathcal{A}$-matrix generated by \texttt{FindAMatrix} using TOPCOM.
Below, we provide details about the input arguments for \texttt{FindTriangulations}.
\begin{itemize}
    \item \texttt{FindAMatrixOut}: Result returned by \texttt{FindAMatrix}.
    \item Options:
    \begin{itemize}
        \item \texttt{MaxRegularTriangs}: This option can take either of the two values: \texttt{All}, or some positive integer. Its default value is \texttt{All}. When set to \texttt{All}, \texttt{FindTriangulations} tries to find all possible regular triangulations. When set to a positive integer \textit{n}, \texttt{FindTriangulations} stops after obtaining upto \textit{n} regular triangulations and then scans for unimodular regular triangulations.
        \item \texttt{PrintRegularTriangs}: This option accepts a boolean value. By default, this option is set to \texttt{False}. When set to \texttt{True}, \texttt{FindTriangulations} prints the regular triangulations in addition to the unimodular regular triangulations. 
        \item \texttt{RunInParallel}: This option accepts a boolean value. By default, this option is set to \texttt{False}. When set to \texttt{True}, Mathematica's native parallelization features are used. 
        \item \texttt{Docker}: This option accepts a boolean value.  By default, this option is set to \texttt{False}. When set to \texttt{True}, one can call \texttt{TOPCOM} from a Docker container having the name \texttt{topcom}, to obtain the regular triangulations. This option might come in handy, should one fail to install \texttt{TOPCOM} natively on their system.
    \end{itemize}
\end{itemize}
}
Upon a successful run with all options set to their default values, \texttt{FindTriangulations} prints all the unimodular regular triangulations, and the overall time consumed at this stage of the computation.
\mybox{
\textbf{\texttt{FindInitialIdeals[FindAMatrixOut]}}
\\\\
This external module generates all possible square-free initial ideals corresponding to the $\mathcal{A}$-matrix generated by \texttt{FindAMatrix}.
Below, we provide details about its input arguments.
\begin{itemize}
    \item \texttt{FindAMatrixOut}: Result returned by \texttt{FindAMatrix}.
\end{itemize}
}
Executing \texttt{FindInitialIdeals} prints all possible square-free initial ideals associated with the given $\mathcal{A}$-matrix, and the overall time consumed. This module does the following:
\begin{itemize}
    \item Use \texttt{Macaulay2} to obtain the toric ideals associated with the $\mathcal{A}$-matrix.
    \item Use the \texttt{gfanInterface} package of \texttt{Macaulay2} to find all possible initial ideals.
    \item Filter for all possible square-free initial ideals.
\end{itemize}
We now construct the series solutions for a given choice of unimodular regular triangulation or square-free initial ideal generated by \texttt{FindTriangulations} or \texttt{FindInitialIdeals}, respectively.
\mybox{
\textbf{\texttt{SeriesRepresentation[Triangulations, TriangNum]}}
\\\\
or,
\\\\
\textbf{\texttt{SeriesRepresentation[Ideals, IdealNum]}}
\\\\
where, the input arguments are as follows:
\begin{itemize}
    \item \texttt{Triangulations}: Result returned by \texttt{FindTriangulations}.
    \item \texttt{TriangNum}: Serial number of the unimodular regular triangulation for which the corresponding series solution is to be derived 
    \item \texttt{Ideals}: result returned by \texttt{FindInitialIdeals}.
    \item \texttt{IdealNum}: Serial number of the square-free initial ideal for which the corresponding series solution is to be derived.
    \item Options:
    \begin{itemize}
        \item \texttt{SubstituteScales}: This option accepts a boolean value.
        By default, this option is set to \texttt{True}.
        When set to \texttt{True}, all the generic coefficients $z_j$, are substituted with their appropriate kinematic counterparts while constructing the series solution for the given non-generic Feynman integral. When \texttt{False}, this substitution is not done.
        \item \texttt{ParameterValue}: This option accepts a list of parameter substitutions.
        \\
        By default, this option is assigned an empty list.
    \end{itemize}
\end{itemize}
}
Running \texttt{SeriesRepresentation} with the result returned from \texttt{FindTriangulation} or \texttt{FindInitialIdeals} will
first print the triangulation number \texttt{TriangNum} or the ideal number \texttt{IdealNum}, respectively. It will then print
the number of summation variables involved in the series solution, the scale substitutions done internally to construct the non-generic limit result for \texttt{SubstituteScales $\rightarrow$ True}. For \texttt{SubstituteScales $\rightarrow$ False}, it simply prints the generic scales in terms of $z_j$. Finally, it prints the summands of the $\Gamma$-series terms constituting the series solution and the total time taken.

Subsequently, one may wish to check if the solution returned from \texttt{SeriesRepresentation} can be written in terms of known hypergeometric functions. This functionality has been implemented in the external module \texttt{GetClosedForm}. For single-fold series, it relies on \textit{Mathematica}'s native capabilities to obtain a closed-form expression. However, for double-sums, these capabilities are usually not sufficient, and therefore, we use the package \texttt{Olsson.wl} \cite{Ananthanarayan:2021yar} to try to obtain a closed-form expression for the series solution. Below, we specify the syntax of this module:
\mybox{
\textbf{\texttt{GetClosedForm[MySeries]}}
\\\\
which has the following input arguments:
\begin{itemize}
    \item \texttt{MySeries}: Result returned by \texttt{SeriesRepresentation} or \texttt{GroebnerDeformation}.
    \item Options:
    \begin{itemize}
        \item \texttt{ParameterValue}: This option accepts a list of parameter substitutions.
        \\
        By default, this option is assigned an empty list.
    \end{itemize}
\end{itemize}
}
This module prints closed-form expressions if found, else it prints the original series, in the same order as in \texttt{SeriesRepresentation}.

Beyond double-sums, it is not straightforward to write $\Gamma$-series terms as known hypergeometric functions. Therefore, in the present version \texttt{GetClosedForm} can yield closed form result upto two-fold series. In fact, even in the two-fold case, one might not necessarily be always able to express these terms as hypergeometric closed-forms. 

The final module \texttt{NumericalSum} is meant for numerical summation of the series solution we obtain from \texttt{SeriesRepresentation}. It also works for the series solution derived from the module \texttt{GroebnerDeformation} which is described in appendix \hyperlink{GD-Module}{A.1}.
\mybox{
\textbf{\texttt{NumericalSum[MySeries, SubstitutionRules, SumLim]}}
\\\\
with the following input arguments:
\begin{itemize}
    \item \texttt{MySeries}: Result returned by \texttt{SeriesRepresentation} or \texttt{GroebnerDeformation}. 
    \item \texttt{SubstitutionRules}: This a list to substitute numerical values to parameters and scales of \texttt{MySeries} before numerical summation.
    \item \texttt{SumLim}: The upper-limit for summing over the summation variables. The same upper-limit is mainatined for each summation variable.
    \item Options:
    \begin{itemize}
        \item \texttt{NumericalPrecision}: Can be used for arbitrary precision calculations. By default, this option is set to \texttt{MachinePrecision}.
        \item \texttt{RunInParallel}: This option accepts a boolean value. Its default value is \texttt{False}. When set to \texttt{True} for cases with a rather complicated \texttt{MySeries} or a large \texttt{SumLim}, running \texttt{NumericalSum} may achieve a non-trivial gain in speed.
    \end{itemize}
\end{itemize}
}
Running \texttt{NumericalSum} for a given \texttt{MySeries}, \texttt{SubstitutionRules} and \texttt{SumLim} would print the numerical result, along with the net time taken for the module's execution.

\section{Example : Two-Mass Bubble} \label{sec:Example}
\begin{figure}[htpb!] 
    \centering 
\includegraphics[scale=0.4]{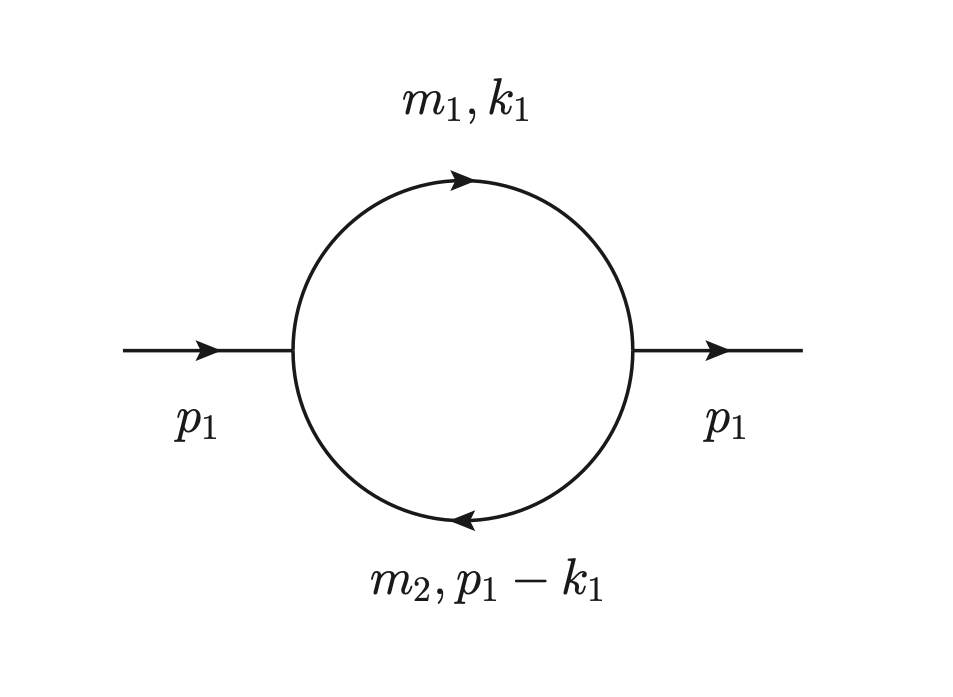} 
    \caption{The Two-Mass bubble diagram} 
\end{figure}

In this section, we provide an explicit demonstration of the typical Feynman integral evaluation procedure using \texttt{FeynGKZ.wl}. We consider the Feynman integral corresponding to the two-mass bubble diagram,

\begin{equation}\label{Bubble_MomRep}
    I_\Gamma(\nu_1,\nu_2,D; p_1^2)= \int \frac{d^Dk_1}{i\pi^\frac{D}{2}}\frac{1}{(-k_1^2+m_1^2)^{\nu_1}(-(p_1+k_1)^2+m_2^2)^{\nu_2}}
\end{equation}
with two unequal masses $m_1$ and $m_2$, and external momentum $p_1$.

We start with specifying the paths to \texttt{FeynGKZ.wl}, the \texttt{polymake} executable and the \texttt{points2triangs} executable from \texttt{TOPCOM}. They are stored as strings in the predefined variables \texttt{FeynGKZPath}, \texttt{PolyMakePath} and \texttt{TOPCOMPath}, respectively. Internally, these variables have been set to the default values \texttt{NotebookDirectory[]}, \texttt{"/usr/bin"} and \texttt{"/usr/local/bin"}, respectively. For our case here, we shall assume that they have been installed in the following non-standard locations.
\mybox{
\inbox{1}{& {\tt FeynGKZPath="/home/user/gkz/math/"}};\\
\gibspace{}{& {\tt PolyMakePath="/home/user/polymake/bin"}};\\
\gibspace{}{& {\tt TOPCOMPath="/home/user/topcom/bin"}};
}
Once this has been done, one can load the package as follows.
\mybox{
\inbox{2}{& {\tt Get[FeynGKZPath<>"FeynGKZ.wl"]}};\\
\boxsplit\\\\
\outbox{2}{& {\tt \textbf{FeynGKZ 1.0} - a Mathematica package for solving}\\
& {$\quad$ {\tt Feynman Integrals using GKZ hypergeometric systems}}}\\
\gibspace{}{& {\tt \textbf{Authors: B. Ananthanarayan, Sumit Banik, Souvik Bera, Sudeepan Datta}}}\\
\gibspace{}{& {\tt \textbf{Last updated:} ${\tt 2^{nd}}$ November, 2022}}
}
Then we specify the momentum representation of the Feynman integral in Eq.\eqref{Bubble_MomRep} as follows,
\mybox{
\inbox{3}{& ${\tt MomentumRep = \{ \{k_1,m_1,a_1\}, \{p_1+k_1,m_2,a_2 \} \}; }$}\\
\gibspace{}{& ${\tt LoopMomenta = \{k_1\};}$}\\
\gibspace{}{& ${\tt InvariantList = \{p_1^2 \rightarrow -s\};}$}\\
\gibspace{}{& ${\tt Dim = 4-2\epsilon;}$}\\
\gibspace{}{& ${\tt Prefactor = 1;}$}
}
Next, we use the module \texttt{FindAMatrix} to compute the $\mathcal{A}$-matrix associated with the GKZ system satisfied by this Feynman integral.
\mybox{
\inbox{4}{& ${\tt FindAMatrixOut = FindAMatrix[\{MomentumRep, LoopMomenta,}$\\
& ${\tt\ InvariantList, Dim, Prefactor \}, UseMB \rightarrow False]; }$}\\
\boxsplit\\
\outbox{2}{& {\tt The Symanzik polynomials $\rightarrow {\tt U = x_1+x_2}$}\\
& ${\tt\ \quad, F = m_1^2x_1^2+sx_1x_2+m_1^2x_1x_2+m_2^2x_1x_2+m_2^2x_2^2 }$}\\
\gibspace{}{& {\tt The Lee-Pomeransky polynomial $\rightarrow {\tt G = }$}\\
& ${\tt\ \quad x_1+m_1^2x_1^2+x_2+sx_1x_2+m_1^2x_1x_2+m_2^2x_1x_2+m_2^2x_2^2}$}\\
\gibspace{}{& {\tt The associated $\mathcal{A}{\tt-matrix}$ $\rightarrow$ $\; \begin{pmatrix}
1 & 1 & 1 & 1 & 1 \\
2 & 1 & 1 & 0 & 0 \\
0 & 1 & 0 & 2 & 1
\end{pmatrix}$, which has ${\tt codim} = {\tt 2}$.}}\\
\gibspace{}{& {\tt Normalized Volume of the associated Newton Polytope $\rightarrow {\tt 3}$}}\\
\gibspace{}{& {\tt Time Taken 1.50005 seconds}}
}
The codimension (\texttt{codim}) of $\mathcal{A}$ is defined as the number of columns minus the number of rows.

After this, we can calculate the unimodular regular triangulations of the point configuration defined by the $\mathcal{A}$-matrix using \texttt{FindTriangulations} as follows:
\mybox{
\inbox{5}{& {\tt Triangulations = FindTriangulations[FindAMatrixOut];}}\\
\boxsplit\\
\outbox{2}{& {\tt Finding all regular triangulations ...}}\\
\gibspace{}{& {\tt Found 5 Regular Triangulations, out of which 3 are Unimodular}}\\
\gibspace{}{& {\tt The 3 Unimodular Regular Triangulations $\rightarrow$}}\\
\gibspace{}{& {\tt \textbf{1 ::} \{\{1,2,3\},\{2,3,4\},\{3,4,5\}\}}}\\
\gibspace{}{& {\tt \textbf{2 ::} \{\{1,2,3\},\{2,4,5\},\{2,3,5\}\}}}\\
\gibspace{}{& {\tt \textbf{3 ::} \{\{2,4,5\},\{1,3,5\},\{1,2,5\}\}}}\\
\gibspace{}{& {\tt Time Taken 0.126965 seconds}}
}
Alternatively, we can also calculate all possible square-free initial ideals from the $\mathcal{A}$-matrix using \texttt{FindIntialIdeals},
\mybox{
\inbox{6}{& {\tt InitialIdeals = FindInitialIdeals[FindAMatrixOut];}}\\
\boxsplit\\
\outbox{2}{& {\tt There are 3 possible square-free initial ideals :}}\\
\gibspace{}{& {\tt \textbf{1 ::} \{$\mathrm{t_2 t_5}$, $\mathrm{t_1 t_5}$, $\mathrm{t_1 t_4}$\}}}\\
\gibspace{}{& {\tt \textbf{2 ::} \{$\mathrm{t_1 t_5}$, $\mathrm{t_1 t_4}$, $\mathrm{t_3 t_4}$\}}}\\
\gibspace{}{& {\tt \textbf{3 ::} \{$\mathrm{t_1 t_4}$, $\mathrm{t_3 t_4}$, $\mathrm{t_2 t_3}$\}}}\\
\gibspace{}{& {\tt Time Taken 3.38268 seconds}}
}

Next, one can find the $\Gamma$-series solutions for the non-generic Feynman integral either for a given choice of square-free initial ideal or unimodular regular triangulation. As described earlier in the documentation, this is done using \texttt{SeriesRepresentation}. We derive the series solution corresponding to the second unimodular regular triangulation as follows,
\mybox{
\inbox{7}{& {\tt SeriesSolution = SeriesRepresentation[Triangulations,2];}}\\
\boxsplit\\
\outbox{2}{& {\tt Unimodular Triangulation $\rightarrow$ 2}}\\
\gibspace{}{& {\tt Number of summation variables $\rightarrow$ 2}}\\
\gibspace{}{& {\tt Non-generic limit $\rightarrow$ \{$ \mathit{z}_{\tt 1}\rightarrow {\tt m_1^2}$, $ \mathit{z}_{\tt 2}\rightarrow {\tt s+m_1^2+m_2^2}$, $ \mathit{z}_{\tt 3}\rightarrow {\tt 1}$, $ \mathit{z}_{\tt 4}\rightarrow {\tt m_2^2}$, $ \mathit{z}_{\tt 5}\rightarrow {\tt 1}$\}}}\\
\gibspace{}{& {\tt The series solution is the sum of following 3 terms.}}\\
\gibspace{}{& {\tt \textbf{Term 1} ::}}\\
\gibspace{}{&  $\biggl(\biggl({\tt (-1)^{-n_1-n_2}\; Gamma[-2+\epsilon+a_1-n_1-n_2]\; Gamma[4-2\epsilon-a_1-a_2+n_2]}$\\
& ${\tt\ \quad \; Gamma[a_2+2n_1+n_2]\; (m_1^2)^{2-\epsilon-a_1} \biggl(\dfrac{m_1^2m_2^2}{(s+m_1^2+m_2^2)^2}\biggl)^{n_1} \biggl(\dfrac{m_1^2}{s+m_1^2+m_2^2}\biggl)^{n_2}}$\\
& ${\tt \quad \; (s+m_1^2+m_2^2)^{-a_2}\biggl)\biggl/\bigl(Gamma[a_1]\;Gamma[4-2\epsilon-a_1-a_2]\; Gamma[a_2]\;}$\\
& ${\tt \quad \; Gamma[1+n_1]\; Gamma[1+n_2]\bigl)\biggl)}$}\\
\gibspace{}{& {\tt \textbf{Term 2} ::}}\\
\gibspace{}{& $\biggl(\biggl({\tt (-1)^{-n_1-n_2}\;
Gamma[-2+\epsilon+a_2-n_1-n_2]\; Gamma[4-2\epsilon-a_1-a_2+n_2]}$\\
& ${\tt \quad \; Gamma[a_1+2n_1+n_2]\;(m_2^2)^{2-\epsilon-a_2}\biggl(\dfrac{m_1^2m_2^2}{(s+m_1^2+m_2^2)^2}\biggl)^{n_1}\biggl(\dfrac{m_2^2}{s+m_1^2+m_2^2}\biggl)^{n_2}}$\\
& ${\tt \quad \; (s+m_1^2+m_2^2)^{-a_1}\biggl)\biggl/\bigl(Gamma[a_1]\;Gamma[4-2\epsilon-a_1-a_2]\;Gamma[a_2]}$\\
& ${\tt \quad \;Gamma[1+n_1]\;Gamma[1+n_2]\bigl)\biggl)}$}
\gibspace{}{& {\tt \textbf{Term 3} ::}}\\
\gibspace{}{& $\biggl(\biggl({\tt (-1)^{-n_1-n_2}
Gamma[2-\epsilon-a_2+n_1-n_2]\;Gamma[2-\epsilon-a_1-n_1+n_2]}$\\
& ${\tt \quad \; Gamma[-2+\epsilon+a_1+a_2+n_1+n_2]\;\biggl(\dfrac{m_1^2}{s+m_1^2+m_2^2}\biggl)^{n_1}\;\biggl(\dfrac{m_2^2}{s+m_1^2+m_2^2}\biggl)^{n_2}}$\\
& ${\tt \quad \; (s+m_1^2+m_2^2)^{2-\epsilon-a_1-a_2}\biggl)\biggl/ \bigl(Gamma[a_1]\;Gamma[4-2\epsilon-a_1-a_2]}$\\
& ${\tt \quad \; Gamma[a_2]\;Gamma[1+n_1]\;Gamma[1+n_2]\bigl)\biggl)}$}\\
\gibspace{}{& {\tt Time Taken 0.066558 seconds}}
}
where $n_1$ and $n_2$ are the summation indices, each of which runs from $0$ to $\infty$. Let us now try to express these results in terms of known hypergeometric functions using the module \texttt{GetClosedForm} which internally uses \texttt{Olsson.wl} \cite{Ananthanarayan:2021yar} for this step.
\mybox{
\inbox{8}{& {\tt GetClosedForm[SeriesSolution];}}\\
\boxsplit\\
\outbox{2}{& {\tt Closed form found with Olsson!}}\\
\gibspace{}{& {\tt \textbf{Term 1} ::}}\\
\gibspace{}{& ${\tt \dfrac{1}{Gamma[a_1]}\;Gamma[-2+\epsilon+a_1]}$\\
& ${\tt \quad \; H3\Big[a_2,4-2\epsilon-a_1-a_2,3-\epsilon-a_1,\dfrac{m_1^2m_2^2}{(s+m_1^2+m_2^2)^2},\dfrac{m_1^2}{s+m_1^2+m_2^2}\Big]}$\\
& ${\tt \quad \; m_1^4\;(m_1^2)^{-\epsilon-a_1}(s+m_1^2+m_2^2)^{-a_2}}$}\\
\gibspace{}{& {\tt \textbf{Term 2} ::}}\\
\gibspace{}{& ${\tt \dfrac{1}{Gamma[a_2]}\;Gamma[-2+\epsilon+a_2]}$\\
& ${\tt \quad \; H3\Big[a_1,4-2\epsilon-a_1-a_2,3-\epsilon-a_2,\dfrac{m_1^2m_2^2}{(s+m_1^2+m_2^2)^2},\dfrac{m_2^2}{s+m_1^2+m_2^2}\Big]}$\\
& ${\tt \quad \; m_2^4\;(m_2^2)^{-\epsilon-a_2}(s+m_1^2+m_2^2)^{-a_1}}$}\\
\gibspace{}{& {\tt \textbf{Term 3} ::}}\\
\gibspace{}{& ${\tt \bigg(\bigg(G1\Big[-2+\epsilon+a_1+a_2,2-\epsilon-a_1,2-\epsilon-a_2,-\dfrac{m_2^2}{s+m_1^2+m_2^2}}$\\
& ${\tt \quad \; ,-\dfrac{m_1^2}{s+m_1^2+m_2^2}\Big]\;Gamma[2-\epsilon-a_1]\;Gamma[2-\epsilon-a_2]}$\\
& ${\tt \quad \; Gamma[-2+\epsilon+a_1+a_2]\;(s+m_1^2+m_2^2)^{2-\epsilon-a_1-a_2}\bigg)\bigg/\big(Gamma[a_1]}$\\
& ${\tt \quad \; Gamma[4-2\epsilon-a_1-a_2]\;Gamma[a_2]\big)\bigg)}$}\\
\gibspace{}{& {\tt Time Taken 0.05827 seconds}}
}
where, \texttt{H3} and \texttt{G1} are the double variable Horn $H_3$ and $G_1$ functions \cite{Srivastava:1985}.
\\
We can also numerically evaluate the result returned in \texttt{Out[7]}, which has been stored in the variable \texttt{SeriesSolution}. 
\mybox{
\inbox{9}{& ${\tt SumLim=30;}$}\\
\gibspace{}{& ${\tt ParameterSub=\{\epsilon \rightarrow 0.001,a_1\rightarrow 1,a_2\rightarrow 1,s\rightarrow 10,m_1\rightarrow 0.4,m_2\rightarrow 0.3\};}$}\\
\gibspace{}{& ${\tt NumericalSum[SeriesSolution,\: ParameterSub,\: SumLim];}$}\\
\boxsplit\\
\outbox{2}{& {\tt Numerical result = 997.382}}\\
\gibspace{}{& {\tt Time Taken 0.222572 seconds}}
}

\section{ Discussions and Conclusions}\label{sec:Conclusion}

In this work, we have considered two very active and interconnected fields of research: Feynman integrals and multivariate hypergeometric functions. While the former is of great importance to the particle physics community as they present challenges at the frontier with consequences to precision tests of the Standard Model, the latter is a rich field in mathematics which has been studied for over a century and found numerous applications in several domains of physics. The connection between these two diverse fields of research was first pointed out in late 1960s by Regge \cite{Regge:1968rhi}. Since then, considerable efforts have been made to study them in parallel, and our work in this paper is motivated by these considerations.

One systematic approach to study multivariate hypergeometric functions came from GKZ, in the form of what we know today as the theory of GKZ hypergeometric systems. Multivariate hypergeometric series are solutions to the partial differential equations of the GKZ system. The mechanism to find a basis of solutions for the GKZ system was advanced by the original authors themselves in the triangulation scheme. On the other hand, SST introduced the GD method of finding solutions to a GKZ system, which is inspired by the Frobenius method to solving ordinary differential equations.  Its connection to Feynman integrals has become explicit in the recent years due to development of the LP representation for counting the number of master integrals for any given family of Feynman integrals \cite{Lee:2013hzt}. 

To recall, the GD and the triangulation methods, were recently used to evaluate Feynman integrals in  \cite{delaCruz:2019skx} and \cite{Klausen:2019hrg} respectively. In both the aforementioned articles, the LP polynomial is used to construct the $\mathcal{A}$-matrix.  Alternatively, one can start from the MB representation and derive the associated GKZ system for a given Feynman integral, as advocated in \cite{Kalmykov:2012rr,Feng:2019bdx,Feng:2022ude}. Although, the methods were algorithmic, no explicit computer programs were publicly available.

We have presented in this work a working code automating the above methods for solving GKZ systems that are associated with Feynman integrals. Our package is based on \textit{Mathematica}, which relies on \texttt{Macaulay2}, \texttt{polymake }and \texttt{TOPCOM}, for the GD and triangulation methods, respectively.  We have described in detail these methods and their implementation with adequate usage documentation, and have also provided our code in the public domain. We believe that our implementation would also help the user to appreciate the underlying equivalence of both the methods, particularly in the setting of Feynman integral evaluation. To this end, we have maintained the following principles in our implementation:
\begin{itemize}
    \item Enabling sufficient flexibility in the usage, so that the user can take an informed decision about the method they would feel at ease to proceed with in a particular circumstance - which would also help them in ensuring that the results obtained from both the methods are always indeed the same.
    \item Provision of two commands that could be used to explicitly test the equivalence  of unimodular regular triangulations and square-free initial ideals. The details of these commands have been provided in appendix \hyperlink{TriIdealCorrespondence}{A.2}.
\end{itemize}
We have also automated the part where we can obtain the $\mathcal{A}$-matrix from the MB representation. Our algorithm is inspired by considerations in \cite{Feng:2019bdx} and its rigorous validity is a subject of ongoing work. 

We have tested our package analytically as well as numerically using \texttt{FIESTA} \cite{Smirnov:2021rhf} by evaluating several familiar yet reasonably non-trivial Feynman integrals for which results mostly exist in the literature. One of them includes the non-trivial two-loop self-energy with four propagators presented recently in \cite{Feng:2022ude}. Several other examples can be found in the \textit{Mathematica} notebook \texttt{Examples.nb} which can downloaded along with the package from \href{https://github.com/anant-group/FeynGKZ}{\texttt{GitHub}}. 

It may be noted here, the time taken by the command \texttt{FindTriangulations} to find all the unimodular regular triangulations is typically lesser than the same for the command \texttt{FindInitialIdeals} that is used to find all the square-free initial ideals. In this sense, the command \texttt{points2triangs} of \texttt{TOPCOM} works faster than the \texttt{gfan} command of \texttt{Macaulay2}. There is one more advantage of the triangulation method. The process of finding the unimodular regular triangulations can be stopped midway, after yielding the required number of triangulations. This is not possible in the \texttt{gfan} command of \texttt{Macaulay2}. Even so, the results coming from the GD method provide a useful crosscheck of the results obtained through the triangulation method.

\texttt{FeynGKZ.wl} yields several series solutions in terms of multivariate hypergeometric functions that are analytic continuations of each other, each being valid in its own convergence region.
These solutions are valid for generic values of the dimension and propagator powers. The evaluation of these solutions for integer powers of propagators where logarithmic terms appear, could be carried out by performing a suitable limit-taking procedure manually. We hope to automate this in a future version of the package. This is however not a restriction in the MB approach \cite{Ananthanarayan:2020fhl}, where one uses multivariate residues to evaluate the integer propagator case and obtain logarithmic series solutions. Therefore, in future, it will be useful to study a cross-fertilization of the two techniques.

\section*{Acknowledgements}

B. Ananthanarayan thanks the Albert Einstein Centre for Fundamental Physics at the University of Bern, Switzerland, for hospitality at the time this manuscript was being prepared. B. Ananthanarayan and Sumit Banik thank the organizers of \textit{Tropical and Convex Geometry and Feynman integrals} at the ETH, Zurich, for inviting them to participate, where many exciting discussions on related topics took place. Sumit Banik thanks the organizers 
of \textit{Elliptic Integrals in Fundamental Physics} at MITP, Mainz, for an invitation to participate, where related discussions took place. Sumit Banik thanks V. Chestnov, R. P. Klausen, H. Munch and F. Tellander for enriching conversations on GKZ theory.
\hypertarget{ExtModules}{\section*{A \quad Other external modules - Documentation} \label{app:ExternalModules}}
In this section, we provide the documentation for the three remaining external modules that were not covered in the main text.

\hypertarget{GD-Module}{\subsection*{A.1 \quad The \texttt{GroebnerDeformation} external module}}

This external module provides an alternative method to construct the $\Gamma$-series solutions, employing the SST algorithm. Its syntax is as follows:
\mybox{\textbf{\texttt{GroebnerDeformation[FindAMatrixOut]}}\\\\
with the following input arguments:
\begin{itemize}
    \item \texttt{FindAMatrixOut}: result returned by \texttt{FindAMatrix}.
    \item Options:
    \begin{itemize}
        \item \texttt{InitialIdeal}: Serial number of the square-free initial ideal for which the corresponding series solution is to be derived. Its default value is \texttt{False}.
        \item \texttt{Weight}: This option accepts a list of non-negative integers, whose length is same as the number of columns of the $\mathcal{A}$-matrix. Its default value is \texttt{False}.
        \item \texttt{AllInitialIdeals}: This option accepts a boolean value. Its default value is \texttt{False}. When set to \texttt{True}, this module yields series solutions for all the possible square-free initial ideals sequentially.
        \item \texttt{ParameterValue}: This option accepts a list of parameter substitutions. By default, this option is assigned an empty list.
    \end{itemize}
\end{itemize}
}
Note: the options \texttt{InitialIdeal} and \texttt{Weight} cannot be used simultaneously. One of them must be set to \texttt{False} while using the other. If the given weight vector is generic, the series solution for the corresponding initial ideal is obtained, otherwise the module prints an error message.

\hypertarget{TriIdealCorrespondence}{\subsection*{A.2 \quad The \texttt{TriangulationToIdeal} and {\texttt{IdealToTriangulation} external modules}}}

These external modules are useful in mapping unimodular regular triangulations to the corresponding square-free initial ideals, and vice-versa. They rely on \texttt{Macaulay2} to function as intended.
\\
Below, we describe the syntax for each of these external modules:
\mybox{
\textbf{\texttt{TriangulationToIdeal[Triangulations, TriangNum]}}
\\\\
where, the input arguments are as follows:
\begin{itemize}
    \item \texttt{Triangulations}: Result returned by \texttt{FindTriangulations}.
    \item \texttt{TriangNum}: Serial number of the unimodular regular triangulation for which the corresponding square-free initial ideal is to be derived.
\end{itemize}
}
and,
\mybox{
\textbf{\texttt{IdealToTriangulation[Ideals, IdealNum]}}
\\\\
where, the input arguments are as follows:
\begin{itemize}
    \item \texttt{Ideals}: Result returned by \texttt{FindInitialIdeals}.
    \item \texttt{IdealNum}: Serial number of the square-free initial ideal for which the corresponding unimodular regular triangulation is to be derived.
\end{itemize}
}
\hypertarget{GlobalVars}{\section*{B \quad Global variables used in \texttt{FeynGKZ.wl}} \label{app:GlobalVar}}
Below, we enlist the global variables that have been used in our package. The user can modify some of them as per their convenience, while others are not meant for such usage.
\mybox{
\begin{itemize}
    \item \textbf{\texttt{FeynGKZPath}}
\\
Set this variable such that it points to the directory 
containing \texttt{FeynGKZ.wl} and its dependencies \texttt{AMBREv2.1.1.m} and \texttt{Olsson.wl}.
\\
Note: the path must be passed as a string.
\\
By default, it points to \texttt{NotebookDirectory[]}.
    \item \textbf{\texttt{PolyMakePath}}
\\
Set this variable such that it points to the directory 
containing the \texttt{polymake} executable.
\\
Note: the path must be passed as a string.
\\
By default, it points to \texttt{/usr/bin}.
    \item \textbf{\texttt{TOPCOMPath}}
\\
Set this variable such that it points to the directory 
containing the \texttt{points2triangs} executable from \texttt{TOPCOM}.
\\
Note: the path must be passed as a string.
\\
By default, it points to \texttt{/usr/local/bin}.
    \item \textbf{\texttt{Dim}}
\\
Space-time dimension of the Feynman integral.
    \item \textbf{\texttt{n}}
\\
Summation indices.
    \item $\bm{\mathit{z}}$
\\
Generic coefficients of the toric $G$-polynomial.
\end{itemize}
}

\hypertarget{Examples}{\section*{C \quad Feynman integrals in \texttt{Example.nb}}}

In this appendix, we provide the list of the Feynman integrals solved in \texttt{Examples.nb} using \texttt{FeynGKZ.wl}. The list is organised in terms of the codimension of the associated $\mathcal{A}$-matrices. All the integrals are evaluated in the $d= 4- 2 \epsilon$ dimension.

\begin{enumerate}
    \item \textbf{Codim $=0$}
    \begin{itemize}
        \item Massless Bubble : Here, the massless one-loop bubble with off-shell external momentum is considered. We find the solution as a single term using the triangulation method.
        \item On-Shell Massless Triangle : Here, the one-loop triangle integral with zero internal masses and  external momenta $p_1^2 = 0, p_2^2 = 0$ and $ p_1 \cdot p_2 =-\frac{s}{2}$ is considered. We find the solution as a single term using the triangulation method.
    \end{itemize}
    \item \textbf{Codim $=1$}
    \begin{itemize}
        \item One-Mass Bubble : Here, the one-loop bubble with one massive propagator with off-shell external momentum is considered. The GD method is used to find the series solution.
        \item On-Shell Massless Box : Here, the one-loop box integral with all massless propagators and on-shell external momenta $p_1^2 = p_2^2 =p_3^2= 0$ is considered. The triangulation method is used to find the series solution.
    \end{itemize}
    \item \textbf{Codim $=2$}
    \begin{itemize}
        \item Off-Shell Massless Triangle : Here, the one-loop triangle integral with massless propagators and off-shell external momenta is considered. The GD method is used to find the series solution.
        \item Two-Mass Bubble : Here, the one-loop bubble with two massive propagator with off-shell external momentum is considered. The triangulation method is used to find the series solution.
        \item On-shell massive sunset : Here, the two-loop sunset diagram with three massive propagator and external momentum $p^2 = 0$ is considered. The triangulation method is used to find the series solution.
    \end{itemize}
    \item \textbf{Codim $=3$}
    \begin{itemize}
        \item Three-Mass Sunset : Here, the two-loop sunset diagram with three massive propagator and off-shell external momentum is considered. The triangulation method is used to find the series solution.
    \end{itemize}
    \item \textbf{Codim $=4$}
    \begin{itemize}
        \item Two-Loop Self-Energy with Four Propagators : Here, the two-loop self energy integral with four propagators is considered. The triangulation method is used to find the series solution.
    \end{itemize}
    \item \textbf{Codim $=5$}
    \begin{itemize}
        \item Off-Shell Massive Triangle : Here, the one-loop triangle integral with three massive propagators and off-shell external momenta is considered. The triangulation method is used to find the series solution.
    \end{itemize}
    
\end{enumerate}

These examples were considered in \cite{delaCruz:2019skx,Klausen:2019hrg,Feng:2019bdx,Feng:2022ude} using the GKZ approach. The solutions of these Feynman integrals obtained from our package provide a crosscheck on the results presented in the literature and match numerically with \texttt{FIESTA} \cite{Smirnov:2021rhf}. 
\newpage
\printbibliography

\end{document}